\documentclass[
reprint,
superscriptaddress,
tightenlines,
aps,
prl,
floatfix
]{revtex4-1}

\usepackage{amsmath}
\usepackage{amssymb}
\usepackage{amsthm}
\usepackage{chngcntr}
\usepackage{dcolumn}% Align table columns on decimal point
\usepackage{float}
\usepackage{graphicx} % Required for inserting images
\usepackage[colorlinks=true, allcolors=blue]{hyperref}
\usepackage{lipsum}
\usepackage{tabularray}
\newcommand{\modified}[1]{\textcolor{blue}{#1}}

\begin{document}

\title{Synergistic signatures of group mechanisms in higher-order systems}

\author{Thomas Robiglio}
\email{robiglio\_thomas@phd.ceu.edu}
\affiliation{Department of Network and Data Science, Central European University, Vienna, Austria}
\author{Matteo Neri}
\affiliation{Institut de Neurosciences de la Timone, Aix Marseille Université, UMR 7289 CNRS, 13005, Marseille, France}
\author{Davide Coppes}
\affiliation{Department of Physics, University of Turin, Turin, Italy}
\author{Cosimo Agostinelli}
\affiliation{Department of Physics, University of Turin, Turin, Italy}
\author{Federico Battiston}
\affiliation{Department of Network and Data Science, Central European University, Vienna, Austria}
\author{Maxime Lucas}
\email{maxime.lucas@unamur.be}
\affiliation{CENTAI Institute, Turin, Italy}
\author{Giovanni Petri}
\email{giovanni.petri@nulondon.ac.uk}
\affiliation{NPLab, Network Science Institute, Northeastern University London, London, UK}
\affiliation{CENTAI Institute, Turin, Italy}
\date{\today} 

% Abstract
\begin{abstract} 
The interplay between causal mechanisms and emerging collective behaviors is a central aspect of understanding, controlling, and predicting complex networked systems. 
In our work, we investigate the relationship between higher-order mechanisms and higher-order behavioral observables in two representative models with group interactions: a simplicial Ising model and a social contagion model.
In both systems, we find that group (higher-order) interactions show emergent synergistic (higher-order) behavior.
The emergent synergy appears only at the group level and depends in a complex, non-linear way on the trade-off between the strengths of the low- and higher-order mechanisms and is invisible to low-order behavioral observables. 
Our work sets the basis for systematically investigating the relation between causal mechanisms and behavioral patterns in complex networked systems with group interactions, offering a robust methodological framework to tackle this challenging task.
\end{abstract}
\maketitle

\begin{figure}
    \centering
    \includegraphics[width=\linewidth]{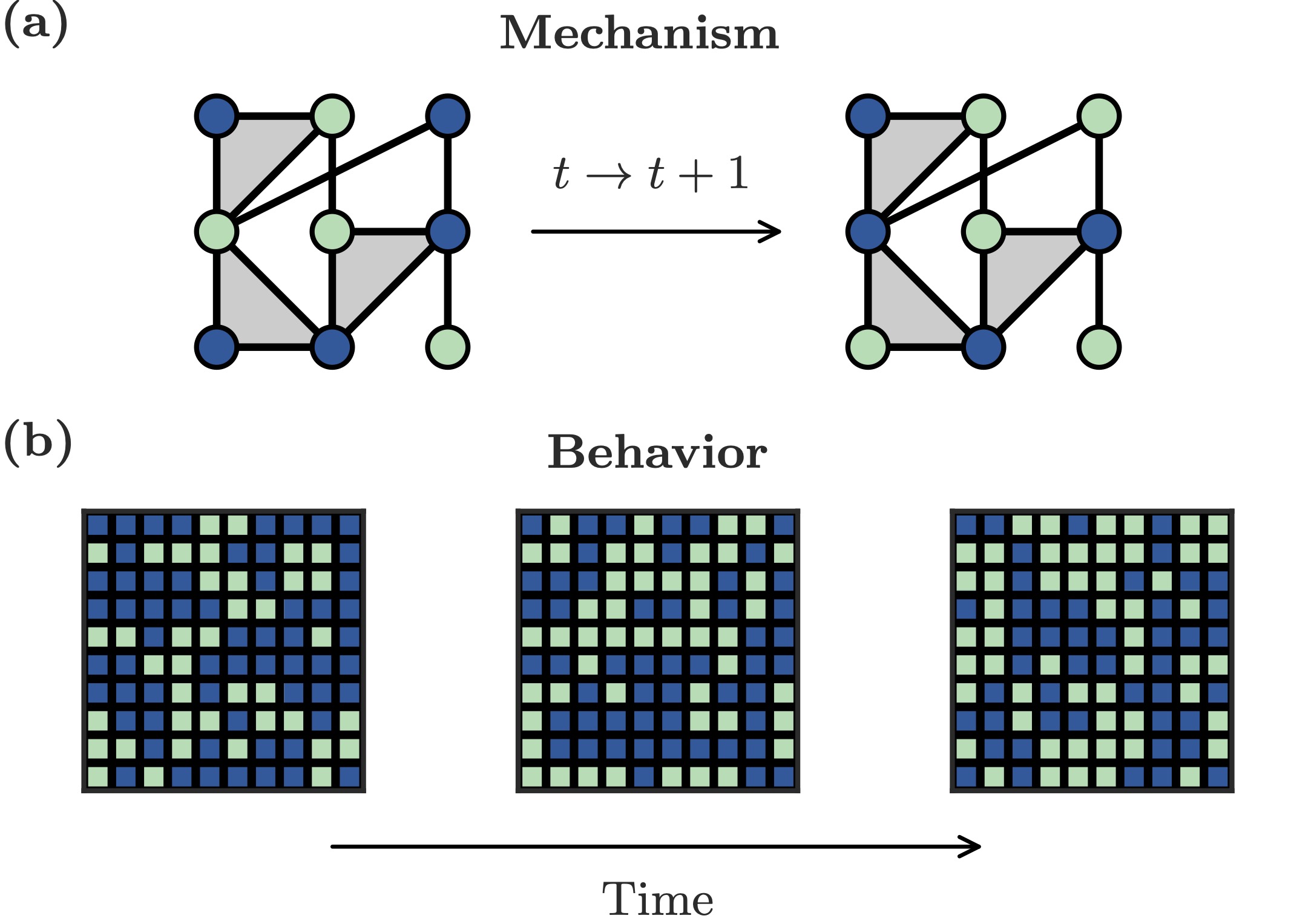}
    \caption{
    \textbf{Mechanisms versus behaviors in complex systems.}
    (a) Mechanisms consist of (i) the topological structure of interactions between nodes and (ii) the rules controlling the temporal evolution of the nodes' states.
    (b) Behaviors are the observable states of the system and encompass its spatial and temporal patterns, interdependencies between units, and emergent phenomena.
    In experimental settings, behaviors are often the only available.
    }
    \label{fig:drawing_mecha_behav}
    \vspace{-1em}
\end{figure}

Mechanisms and behaviors are two facets of the study of complex systems:
\textit{mechanisms} are the structural and dynamical rules controlling the causal evolution of the system; \textit{behaviors}, instead, refer to the measurable observables quantifying statistical interdependencies between units of a system in space and time (Fig. \ref{fig:drawing_mecha_behav}).
The nature of the relation between the two facets and the limits of our capacity to reconstruct it is a long-standing problem in the analysis of complex systems \cite{grenander1994representations,sterman1994learning,scarpino2019predictability,han2015robust,squartini2018reconstruction, peixoto2018reconstructing,prasse2022predicting}.

Existing methods to study each of the two facets mostly adopt lower-order descriptions: pairwise network representations for mechanisms~\cite{barabasi2012network,peixoto2019network}, and low-order information-theoretic metrics for behaviors~\cite{cliff2023unifying,bianco2016successful}.
Despite their success, these low-order methods often fail to fully capture the intricate nuances inherent to many complex systems~\cite{battiston2021physics,rosas2022disentangling}, thus beyond-pairwise methods are being developed:
higher-order network representations such as hypergraph or simplicial complexes~\cite{battiston2020networks} and higher-order behavorial metrics, both topological~\cite{santoro2023higher} and information-theoretic~\cite{rosas2019quantifying}.

A central question is then: \textit{what is the relation between higher-order mechanisms and behaviors?}
The presence of higher-order mechanisms enhances pairwise interdependencies, measurable for instance with mutual information or pairwise correlations.
On the other hand, intuition might suggest that observing higher-order behaviors implies the presence of higher-order mechanisms.
However, this is not the case.  
Systems with only low-order mechanisms can display higher-order behaviors: for example, a simple system of three spins connected by pairwise anti-ferromagnetic interactions shows a total interdependency (higher-order behavior) significantly larger than the sum of the three pairwise interdependencies (low-order behaviors)~\cite{matsuda2000physical, rosas2022disentangling}.
As both low and higher-order mechanisms can determine the observation of both low and higher-order behaviors, the connection between behavioral observables and microscopic mechanisms in systems with pairwise and group interactions is not trivial; a systematic investigation of this complex relationship across different orders of interactions is needed \cite{neri2024taxonomy}.

Here, we explore the mechanism-behavior relation in higher-order versions of two canonical dynamical processes---a generalization of the Ising model~\cite{huang2009introduction, dorogovtsev2008critical}, and a social contagion model~\cite{iacopini2019simplicial}---and quantify higher-order behavior by defining the \textit{total dynamical O-information}, an extension of transfer entropy to arbitrary groups of variables~\cite{rosas2019quantifying, stramaglia2021quantifying}.
In both systems, we uncover an emergent synergistic behavioral signature of group interactions.
Synergistic behaviors manifest when information about a group of variables can only be recovered by considering the joint state of all variables and cannot be reconstructed from subsets of units of the group.
Crucially, the observed behavioral signatures display a complex non-linear dependence on the strength of the higher-order mechanisms. 
While these signatures can, in some regimes, be overshadowed by other emergent phenomena in the systems (\textit{e.g.} the transition to the magnetized phase in the Ising model), when present, they are invisible to low-order observables and thus represent genuine higher-order phenomena.

\paragraph{Quantifying higher-order behaviors.}
The partial information decomposition framework allows for the characterization of the information-sharing interdependencies between groups of variables~\cite{williams2010nonnegative, griffith2014quantifying, varley2023partial}. 
Qualitatively, these relations can be of three types: redundant, synergistic, or unique.
Consider three variables $X_1$, $X_2$ and $X_3$. 
Information is redundant if it is replicated over the variables (\textit{i.e.} recoverable from $X_1 \lor X_2 \lor X_3$), synergistic if it can only be recovered from their joint state ($X_1 \land X_2 \land X_3$), and unique if it can only be recovered from one variable and nowhere else.
In this framework, mutual information has been extended to groups of three or more variables by the O-information~\cite{rosas2019quantifying}.
To generalize the O-information of multivariate time series from equal-time correlations to time-lagged correlations---similarly to how transfer entropy extends mutual information~\cite{schreiber2000measuring}---Stramaglia \textit{et al.} proposed \textit{dynamical} O-information~\cite{stramaglia2021quantifying}.
This quantity is defined by (i) considering $n$ variables $\mathbf{X}=(X_1,...,X_n)$ on which we compute the O-information $\Omega_{n}(\mathbf{X})$, (ii) adding a new variable $Y$, and (iii) computing the variation of O-information: $\Delta_n = \Omega_{n+1}(\mathbf{X},Y) - \Omega_{n}(\mathbf{X})$.
To remove shared information due to common history, the dynamical O-information is defined by conditioning $\Delta_n$ on the history $Y_0$ of the target variable $Y$:
\begin{align}
    \label{eq:dynamical_o_info}
    d\Omega_n(Y;\mathbf{X}) \equiv & (1-n) \, \mathcal{I}(Y;\mathbf{X}|Y_0)\\
    & +\sum_{j=1}^n \mathcal{I}(Y;\mathbf{X}_{-j}|Y_0) \nonumber
\end{align}
Here, $\mathcal{I}(\cdot ; \cdot | \cdot)$ is the conditional mutual information, $Y_0=\left(y(t),y(t-1),...,y(t-\tau+1)\right)$ the past and present of $Y$, and $Y=y(t+1)$ its next instance.
The parameter $\tau$ is the temporal horizon of the time-series, usually set to a relevant time scale of the process.
To quantify the dynamical O-information regardless of source-target assignments, we define the \textit{total} dynamical O-information as:
\begin{equation}
    \label{eq:total_d_o_info}
    d\Omega_n^{\text{tot.}}(\mathbf{X})\equiv\sum_{j=1}^n d\Omega_{n-1}(X_j;\mathbf{X}_{-j}) .
\end{equation}
Total dynamical O-information inherits from O-information the property of being a signed metric: $d\Omega_n^{\rm tot.}(\mathbf{X})>0$ indicates that information-sharing among the units of $\mathbf{X}$ is dominated by redundancy, while $d\Omega_n^{\rm tot.}(\mathbf{X})<0$ indicates that it is dominated by synergy.

\paragraph{Dynamical systems with higher-order mechanisms.}
We consider two discrete higher-order dynamical models: a simplicial Ising model and the simplicial model of social contagion \cite{iacopini2019simplicial}.
Both are defined on simplicial complexes, a class of hypergraphs~\cite{battiston2020networks} that encode multi-node interactions as simplices and respect downward closure.

The first model we consider is a simplicial Ising model.
This model is an extension of the Ising model~\cite{huang2009introduction, dorogovtsev2008critical} with group interactions of different strengths for simplices of different sizes.
We consider a simplicial complex $\mathcal{K}$ with average generalized degrees $\{\langle k_{\ell}\rangle\}$, where each of the $N$ nodes has two possible states: spin-up ($S^i=+1$) or spin-down ($S^i=-1$).
The model is defined by the Hamiltonian:
\begin{align}
\label{eq:simplicial_ising_hamiltonian}
    H = & - J_0 \sum_{i=1}^N S^i + \\
    & - \sum_{\ell=1}^{\ell_{\text{max}}} \frac{J_{\ell}}{\langle k_{\ell}\rangle} 
    \sum_{\{\sigma \in \mathcal{K}:|\sigma|=\ell\}}
    \left[2 \bigotimes_{i\in \sigma}S^i -1 \right] \nonumber
\end{align}
where $\ell_{\text{max}}$ is the maximal order of $\mathcal{K}$ and:
\begin{equation}
    \bigotimes_{i=1}^n S^i =\delta\left(S^1,...,S^n\right)
    =\begin{cases}
    1 & \text{if\;} S^1=...=S^n \\
    0 & \text{otherwise}
    \end{cases}
\end{equation}
is the Kronecker delta for an arbitrary number of binary arguments.
Inserting the Kronecker delta---instead of the product~\cite{merlini1973symmetry, liu2017self, wang2022full}---in the coupling terms is necessary to preserve the symmetry under spin flip at all sites of the dyadic model with \modified{no} magnetic field ($J_0=0$).
We consider the dynamics of this system to be the sequence of Monte Carlo moves performed with the Metropolis-Hastings acceptance-rejection rule~\cite{newman1999monte} at temperature $T$. 

The second model we consider is the simplicial model of social contagion~\cite{iacopini2019simplicial}.
Following the SIS framework~\cite{pastor2015epidemic}, we associate to each of the $N$ nodes of a simplicial complex $\mathcal{K}$ a binary variable $x_i(t)\in \{0,1\}$, corresponding to the susceptible or infected state of agent $i$ at time $t$.
At the initial time step $t_0$, a fraction of infected agents $\rho_0 = \sum_i x_i(t_0)/N$ is placed in the population.
At each time step, susceptible agents ($x_i(t)=0$) become infected with a probability $\beta_{\ell}$ if they belong to a $\ell$-simplex where all other nodes are infected.
Infected agents ($x_i(t)=1$) recover independently with probability $\mu$.
We introduce the usual rescaled infectivity parameter of order $\ell$: $\lambda_{\ell} = \beta_{\ell} \langle k_{\ell}\rangle / \mu$.

For computational feasibility, we limit ourselves to group mechanisms and interdependencies up to three nodes (\textit{i.e.} $\ell_{\text{max}} = 2$). 
The results shown are obtained in random simplicial complexes with $N=200$ nodes and average degrees $\langle k_1 \rangle = 20$, $\langle k_2 \rangle = 6$. 
The Ising model was simulated for $3\times10^4$ time steps, and the contagion model for $10^4$ time steps. 
Other parameters were set to $T=1$ (Ising) and $\mu = 0.8$ and $\rho_0=0.3$ (contagion).

\begin{figure}
    \centering
    \includegraphics[width=\linewidth]{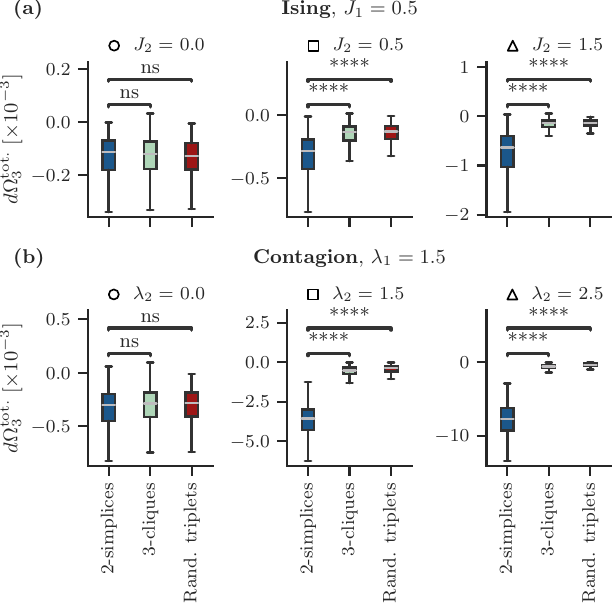}
    \caption{
    \textbf{Synergistic signature of higher-order mechanisms.}
    We show box plots of the distributions of total dynamical O-info $d\Omega_3^{\rm tot.}$ in 
    (a) the simplicial Ising, and
    (b) the simplicial contagion models. 
    Distributions are over all occurrences of three types of groups of nodes: 2-simplices, 3-cliques, and random triplets, and are shown for increasing values of the group mechanism strengths ($J_2$ and $\lambda_2$). As strength increases, higher-order interactions become more synergistic (negative $d\Omega_3^{\rm tot.}$) than lower ones.
    Symbols ``ns" and ``****" indicate a non-significant and significant ($p\leq 10^{-4}$) difference, respectively, between the distributions ($t$-test).}
    \label{fig:boxplots}
    \vspace{-1em}
\end{figure}

\paragraph{Emergence of synergistic signatures of group interactions.}
We simulate the two systems for different values of the control parameters and compute the total dynamical O-information $d\Omega_3^{\text{tot.}}$ on the resulting time-series for different \textit{types} of node triplets: 
2-simplices, 3-cliques, and uniformly randomly chosen triplets of nodes.
The 2-simplices are the true higher-order interactions as the nodes belonging to them will interact through the 3-body ferromagnetic coupling in the Ising model and the group infection rate in the contagion model.
The 3-cliques can be thought of as ``spurious'' higher-order interactions as the nodes belonging to them are all interconnected but through pairwise couplings only.
In both cases, we set the delay of the dynamical O-information to $\tau=1$ as both systems are Markovian.
Increasing the strength of group interactions ($J_2$, the three-body coupling in the Ising model, and $\lambda_2$, the group rescaled infection rate in the contagion model), we observe an increasing co-occurrence of higher-order mechanisms and synergistic higher-order behaviors (Fig.~\ref{fig:boxplots}).
In both systems, all types of triplet groups display synergistic higher-order behaviors ($d\Omega_3^{\text{tot.}}<0$); however, and crucially, as we increase the relative strength of the higher-order mechanisms, we see that 2-simplices, \textit{i.e.} the genuine higher-order interactions, display significantly stronger synergistic behaviors than the other groups.

\begin{figure}
    \centering
    \includegraphics[width=\linewidth]{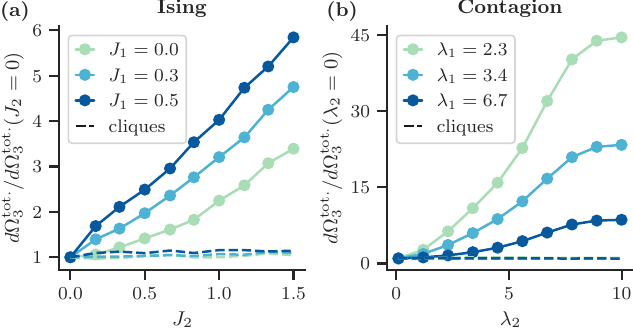}
    \caption{
    \textbf{Complex dependence of higher-order behaviors on higher-order mechanisms.}
    We show the relative variation of total dynamical O-info $d\Omega_3^{\rm tot.}$ measured on 2-simplices as a function of the strength of higher-order mechanisms (a)~the simplicial Ising, and (b)~the
simplicial contagion models. The dashed line shows the same quantity for 3-cliques.
    }
    \label{fig:nonlinear}
    \vspace{-1em}
\end{figure}

\paragraph{Complex dependence of higher-order behaviors on higher-order mechanisms.} To go further, we now show (Fig.~\ref{fig:nonlinear}) how the total dynamical O-info $d\Omega_3^{\rm tot.}$ changes as the strength of group mechanisms is increased ($J_2$ and $\lambda_2$, respectively), relatively to its value without group mechanisms ($J_2=0$ and $\lambda_2=0$, respectively).
We see that, as we increase this group coupling strength, the $d\Omega_3^{\rm tot.}$ measured on 2-simplices (solid lines) shows a relative increase with respect to the case without group mechanisms, regardless of the (color-coded) pairwise coupling strength ($J_1$ and $\lambda_1$, respectively).
Moreover, total dynamical O-info measured on 3-cliques (dashed lines) stays roughly constant as $J_2$ and $\lambda_2$ increase. These two facts confirm the results from Fig. \ref{fig:boxplots}, showing that group mechanisms promote higher-order synergistic behavior. 
More importantly, we see that the relative $d\Omega_3^{\rm tot.}$ response of 2-simplices is qualitatively different in the two systems. First, the response appears to be roughly linear (until the transition to the ferromagnetic phase occurs, see Supplemental Material~\footnote{We present in the Supplemental Material (SM) an introduction to higher-order network structures and the informational theoretical metrics used in this work as well as the computational tools to work with them. In the SM we also present complementary results to the ones shown in the main text.}) in the simplicial Ising model, but non-linear in the simplicial contagion model. 
Second, although the relative $d\Omega_3^{\rm tot.}$ depends on the pairwise coupling strength ($J_1$ and $\lambda_1$, respectively) in both systems, it does so in opposite ways: an increase in the pairwise coupling yields a larger relative $d\Omega_3^{\rm tot.}$ in the simplicial Ising model, but yields a lower relative $d\Omega_3^{\rm tot.}$ in the simplicial contagion model. These results indicate a complex and system-dependent interplay between low- and higher-order mechanisms and behaviors, which requires further investigation.

\paragraph{Insufficiency of lower-order metrics.}
Despite the strong synergistic behaviors displayed by genuine higher-order interactions, we still do not know the extent of this correspondence, nor whether low-order observables could already detect---and to what degree---the presence of higher-order interactions. 
Moreover, we need to determine whether group behaviors are truly higher-order, or the byproduct of low-order interdependencies. 
To answer these questions, we compare our higher-order metric with a lower-order metric over the parameter space of both systems.
For the latter, for each triplet, we compute the sum of the transfer entropies between the time series of the three possible node pairs.
For both metrics, we quantify the difference in behavior between 2-simplices and 3-cliques via the statistical distance $d$~\cite{cover1999elements}~\footnote{Triplets of random nodes behave similarly to 3-cliques, moreover it is of interest to discriminate between true and spurious three-body couplings}. 
For two distributions---here, $P_{2}$ for 2-simplices and $P_{3}$ for 3-cliques---over a common alphabet $\chi$, $d$ is defined as:
\begin{equation}
    \label{eq:stat_dist}
    d (P_{2}, P_{3})=\frac{1}{2}\sum_{x\in \chi} \left|P_{2}(x)-P_{3}(x)\right| ,
\end{equation}
which we denote $d_{23} \equiv d (P_{2}, P_{3})$ for short. 
The distance $d_{23}$ quantifies the overlap of the two distributions. 
By definition, it takes values in  $[0,1]$: $d_{23}=0$ if the two distributions are identical, and $d_{23}=1$ if the two distributions 
take non-zero values on non-overlapping subsets of $\chi$~\footnote{The statistical distance is half of the $L^1$ distance between the two distributions}.

\begin{figure}
    \centering
    \includegraphics[width=\linewidth]{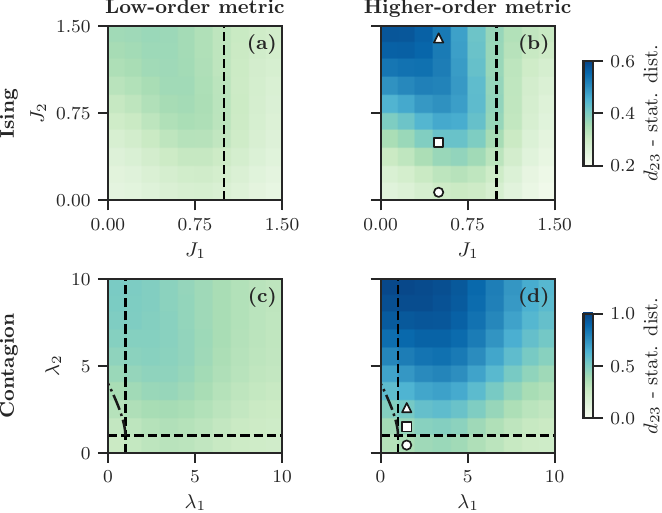}
    \caption{
    \textbf{Low-order metrics do not see the synergistic signature of higher-order mechanisms.}
    We show the statistical distance $d_{23}$ between the distributions of the behavior of 2-simplices and 3-cliques for two metrics: (a), (c) sum of transfer entropies (low-order) and (b), (d) total dynamical O-information (higher-order). Two models are shown. (a), (b) The simplicial Ising model, where the dashed line is the critical coupling strength of the pairwise model $J_1^{\rm cr.}=1$ with no magnetic field. (c), (d) The simplicial contagion model, 
    where the two dashed lines are, respectively, the epidemic threshold of the pairwise SIS model $\lambda_1^{\rm cr.}=1$ and the critical value of the rescaled 2-simplices infectivity rate above which the system shows the discontinuous phase transition and bistability $\lambda_2^{\rm cr.}=1$, and the dash-dotted line represents the points $(\lambda_1,\lambda_2)$ where the system undergoes a discontinuous transition~\cite{iacopini2019simplicial}. The three white symbols in (b) and (d) correspond to the parameter values shown in Fig.~\ref{fig:boxplots}.
    }
    \label{fig:phasespace_dist}
    \vspace{-1em}
\end{figure}

In both systems, we find two main results. 
First, the low-order behavioral metric does not see differences between the lower- (3-cliques) and higher-order mechanisms (2-simplices), whereas the higher-order metric does. 
Indeed, this is indicated by the uniformly low values of $d_{23}$ with low-order metric (Fig.~\ref{fig:phasespace_dist}a,c) with respect to the large values exhibited by $d_{23}$ with the higher-order metric, the total dynamical O-information (Fig.~\ref{fig:phasespace_dist}b,d). 
The latter is consistent with the synergetic signature results shown in Fig. \ref{fig:boxplots}. 
So, the higher-order mechanisms can be identified---and distinguished from low-order mechanisms---by the higher-order behavioral metric and not by the low-order one.

Second, focusing on $d\Omega_3^{\rm tot.}$, we see that the difference between the 2-simplices and 3-cliques is large (large $d_{23}$, dark blue)  over a finite region of parameter space (Fig.~\ref{fig:phasespace_dist}b,d). 
This region corresponds to the co-occurrence of higher-order mechanisms and synergistic higher-order behavior (Fig.~\ref{fig:boxplots}). 
This occurs for sufficiently large values of the strength of the higher-order mechanisms.

These findings apply to both models, but each model has its specificities. 
While explaining the full shape of the dark blue region is a hard task, we can explain some of its features.
In the Ising model (Fig.~\ref{fig:phasespace_dist}b), the large $d_{23}$ $\gtrsim 0.5$ region (dark blue) 
does not extend above $J_1^{\rm cr.}=1$ (dashed line), which is the magnetization threshold of the pairwise model with no magnetic field~\footnote{The Ising model on an Erdős–Rényi graph is well described by a mean-field description~\cite{dorogovtsev2008critical}, which has critical temperature given by $T_{\rm cr.}/J=q$ where $q$ is the coordination number of the lattice. 
Fixing the temperature $T=1$ and with $q=\langle k_1 \rangle$ in our Hamiltonian Eq.~\eqref{eq:simplicial_ising_hamiltonian}, we obtain the critical value of the pairwise coupling: $J_1^{\rm cr.}=1$.
}.
This is because, above that value, the system magnetizes, and no information can be recovered.
Below ($J_1<1$), the region appears above a certain strength of the three-body coupling $J_2\gtrsim 0.5$, and that value seems to decrease as $J_1$ increases.

In the contagion model (Fig.~\ref{fig:phasespace_dist}d) the large $d_{23}$ region displays larger values ($0.7 \lesssim d_{23} \lesssim 1.0$).
It does not appear to be bounded from above. 
However, it is bounded from below. First, for $\lambda_1 < 1$ (left of the vertical dashed line), the region appears only above $\lambda_{\rm cr.} = 2\sqrt{\lambda_2} - \lambda_2$ (dashed-dotted line). 
These values are known: $\lambda_1 = 1$ is the epidemic threshold of the pairwise version of the model (SIS on an Erdős–Rényi graph), and $\lambda_{\rm cr.}$ is the value where the discontinuous transition occurs~\cite{iacopini2019simplicial}. Below that value, only the epidemic-free state exists, for which no higher-order behaviors are expected. 
Second, the region does not appear to extend below $\lambda_2 = 1$ (horizontal dashed line), below which no discontinuous transition can exist, and we thus expect the system to behave more like its low-order variant. 
Finally, for larger $\lambda_1$, the region starts above values of $\lambda_2$ that are larger as $\lambda_1$ increases, suggesting that their ratio plays a role.

In conclusion, by exploring the relation between mechanisms and behaviors in two systems with higher-order interactions, we uncovered emergent synergistic signatures characterizing group mechanisms.
Quantifying higher-order behaviors using $d\Omega_3^{\rm tot.}$, we showed that in both models, an increase in the strength of the parameter controlling the group mechanisms in 2-simplices led (non-linearly) to significantly larger synergistic values of $d\Omega_3^{\rm tot}$.
We have also shown that the synergistic behavioral response of the groups of nodes to the variation of the driving higher-order mechanisms is non-linear and shows system-dependent characteristics.
Crucially, low-order observables did not capture the groups' behavioral signatures, supporting the importance of higher-order observables \modified{to study group interdependencies}.
For example, we present in the Supplemental Material a simple method leveraging these synergistic signatures for the detection of higher-order interactions from the nodes' states' time-series.
By exploring the control parameter spaces of the two systems, we showed that synergistic signatures are not ubiquitous and can be overshadowed by other emergent phenomena (\textit{e.g.} the magnetization transition in the Ising model). 
We expect our results to be relevant for any attempts at reconstructing~\cite{piaggesi2022effective,wang2022full,levina2022tackling,santoro2023higher} and predicting~\cite{murphy2023duality,prasse2022predicting} complex interacting systems from signals, and for the ongoing discussion about the nature and importance of higher-order systems~\cite{thibeault2024low}.

\bibliography{reference}

\end{document}

\title{Supplemental Material}

%%%%%%%%%% Prefix a "S" to all equations, figures, equations, tables and reset the counter %%%%%%%%%%
\renewcommand*{\citenumfont}[1]{S#1}
\renewcommand*{\bibnumfmt}[1]{[S#1]}

\setcounter{figure}{0}
\setcounter{table}{0}
\setcounter{equation}{0}
\setcounter{page}{1}
\setcounter{section}{0}

\makeatletter
\renewcommand{\thepage}{\roman{page}}
\renewcommand{\thefigure}{S\arabic{figure}}
\renewcommand{\theequation}{S\arabic{equation}}
\renewcommand{\thetable}{S\arabic{table}}
%%%%%%%%%% Prefix a "S" to all equations, figures, equations, tables and reset the counter %%%%%%%%%%

\setcounter{secnumdepth}{2}

\maketitle

\onecolumngrid

\section{Higher-order network structures}
\begin{figure}
    \centering
    \includegraphics{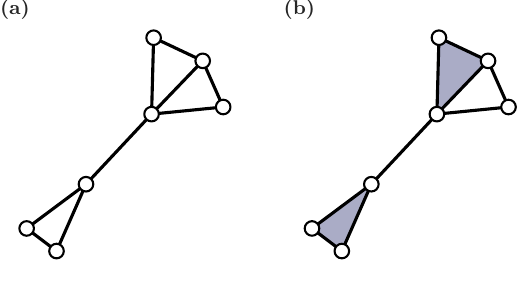}
    \caption{In a system with group interactions when we consider a traditional network representation (a) the group interactions are either projected down or neglected; instead using a higher-order representation (b) such as a simplicial complex allows for the explicit representation of group interactions using simplices.}
    \label{fig:low_vs_high}
\end{figure}

Complex networks theory is the study of phenomena that result from the interplay among many distinct parts interacting in a non-trivial way~\cite{newman2018networks}. 
Examples of such phenomena are everywhere: the spread of fake news on social media, the interaction network of proteins in an organism, the activation of neurons in the brain, climate networks, \textit{etc.}.
Over the past decades, various complex systems have been successfully described as graphs whose interacting pairs of vertices are connected by edges. 
An important limitation of traditional networks, however, is that only pairwise interactions can be represented~\cite{battiston2020networks}. 
This means that the evolution of the system under study can only come from dyads of elements influencing each other. 
In this framework, interactions between groups of agents are typically neglected or projected down as combinations of pairwise interactions (see Fig.~\ref{fig:low_vs_high}).
In recent years, there has been a growing interest of the network science community towards finding and analyzing explicit representations of group interactions, between any number of units in interconnected systems. 

The natural candidates for providing such descriptions are hypergraphs and simplicial complexes.
Hypergraphs are the straightforward generalization of graphs, allowing to encode interactions among arbitrary numbers of nodes.
A hypergraph is defined by a set $V$, whose elements are known as vertices or nodes, and by a family $E$ of subsets of $V$, known as hyperedges. A $k$-hyperedge $\sigma$ is a set of $k+1$ nodes $\sigma = [p_0, p_1,...,p_k]$ with $p_0,...,p_k\in V$.
%A hypergraph is denoted by  $\mathcal{H}=(V,E)$.
Simplicial complexes are a special case of hypergraphs and offer another approach. 
In a simplicial complex, a hyperedge is called a simplex. Contrary to hypergraphs, in a simplicial complex $\mathcal{K}$ all subfaces of a simplex $\sigma$ (for example, the edges of a triangle) need to be included in the simplicial complex. This is called the \textit{inclusion condition} and can be written as:
\begin{equation}
        \sigma \in \mathcal{K} \rightarrow \forall \nu \subset \sigma\text{,\;\;} \nu \in \mathcal{K}
\end{equation}
where $\nu$ is a sub-simplex of $\sigma$.
Although more constrained than hypergraphs, simplicial complexes provide access to powerful mathematical formalisms from algebraic topology~\cite{hatcher2005algebraic}.

\subsection{Random Simplicial Complexes}
The Random Simplicial Complex (RSC) model is a generalization of the Erdos-Renyi model for pairwise networks. The RSC model generates simplicial complexes with simplices of different dimensions, controlling the expected local connectivity~\cite{iacopini2019simplicial}.
The RSC model with simplices up to dimension $D$ has $D+1$ parameters: the number of vertices $N$ and $D$ probabilities $\{p_1,...,p_k,...,p_D\}$, where $p_k\in[0,1]$ controls the probability for the creation of $k$-simplices.
In the case $D=2$ (used in the main text for generating the structures in our analysis) the simplicial complexes are produced as follows. 
Given the $N$ vertices we connect any two nodes $i,j$ with probability $p_1$ (this is equivalent to the $G(N,p)$ ensemble for Erdős-Rényi random graphs). 
Then for any three nodes $l,m,n$ we add a simplex $(l,m,n)$ with probability $p_2$.
At this point, the average generalized degree $\langle k_2 \rangle$ (average number of 2-simplices incident in a node), is given by:
\begin{equation}
\label{eq:k2_RSC}
    \langle k_2 \rangle = \frac{(N-1)(N-2)}{2} p_2
\end{equation}
The structure generated up to now does not satisfy the inclusion condition in general---additional 1-simplices need to be added to ensure that all 2-simplices are ``closed''.  
These new 1-simplices will contribute to increasing the 1-degree of the nodes.

\begin{figure}
    \centering
    \includegraphics{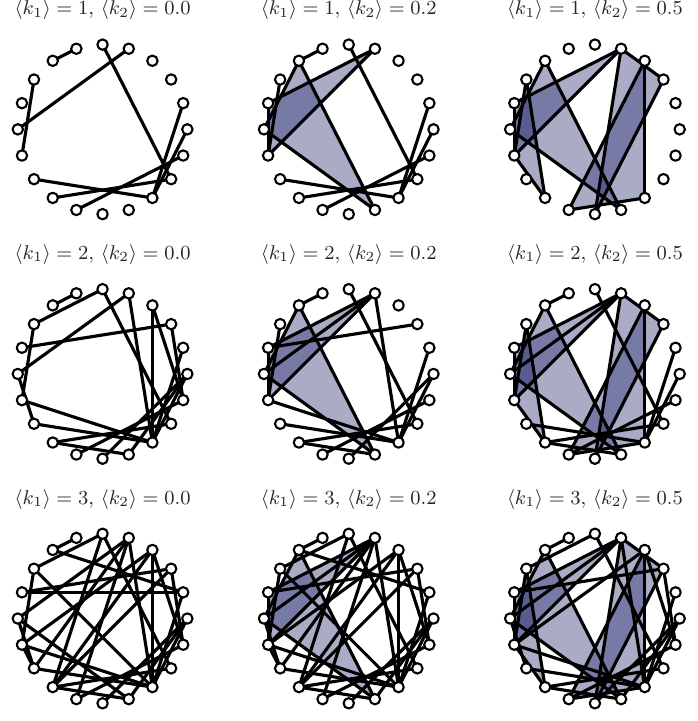}
    \caption{Realizations of the Random Simplicial Complex (RSC) model with $N=20$ and different values of $\langle k_1\rangle$ and $\langle k_2 \rangle$.}
    \label{fig:RSC_p1_p2}
\end{figure}

The average degree $\langle k_1 \rangle$ can now be computed in terms of the probabilities. 
The contribution of the addition of 1-simplices to the expected degree $k_{1,i}$ of a node $i$ can be divided in the following scenarios:
\begin{itemize}
    \item Incremented by 2 for each 2-simplex $(i, j, k)$ such that neither the link $(i, j)$ nor the link $(i, k)$ are already present.
    This occurs with probability $(1-p_1)^2$.
    \item If either the link $(i, j)$ is already present but not $(i, k)$, or vice-versa, the addition of the 2-simplex $(i, j, k)$ increases the degree of $i$ by 1. Since each case happens with the same probability $p_1(1 - p_1)$ the contribution is therefore $2p_1(1 - p_1)$.
\end{itemize}
Thus we can write, for $p_1,p_2 \ll 1$:
\begin{equation}
\label{eq:k1_RSC}
    \langle k_1 \rangle \simeq (N-1)p_1+2\langle k_2 \rangle (1-p_1) 
\end{equation}
We can invert Eqs. \eqref{eq:k2_RSC}-\eqref{eq:k1_RSC} to compute the values of $p_1$ and $p_2$ that results in prescribed values of the average degrees as:
\begin{subequations}
\label{eq:p1p2_from_k1k2}
\begin{align}
    p_1 = & \frac{\langle k_1 \rangle - 2 \langle k_2 \rangle}{N-2\langle k_2 \rangle -1}
    \\
    p_2 = & \frac{2\langle k_2 \rangle}{(N-1)(N-2)}
\end{align}
\end{subequations}
We show in Fig.~\ref{fig:RSC_p1_p2} some examples of RSC with $N=20$ and different values of $\langle k_1\rangle$ and $\langle k_2 \rangle$.
The generation of these structures and their visualization, as well as the rest of the analysis carried out in this paper relies on the XGI Python package~\cite{Landry_XGI_2023}.
This library provides data structures and algorithms for modeling and analyzing complex systems with higher-order interactions.

\section{Information theoretic measures}

In the following section we present the two information theoretic measures that are used in the main text to probe low and higher-order behaviors, namely transfer entropy and total dynamical O-information.

\subsection{Transfer Entropy}
The Transfer Entropy (TE) is an information-theoretic measure of the directed (time-asymmetric) information transfer between two random processes~\cite{schreiber2000measuring}. 
Given two variables $X$ and $Y$, the TE from $X$ to $Y$ is defined as the mutual information between the present of $X$ and the future of $Y$, conditioned on the past of $Y$:
\begin{equation}
\label{eq:Transfer Entropy}
    \mathcal{T}(X\to Y) =
    \mathcal{T}(X;Y)\equiv \mathcal{I}(X_t,Y_{t+1}|Y_{t}) = \mathcal{H}(X_t, Y_{t+1}) + \mathcal{H}(Y_t, Y_{t+1}) - \mathcal{H}(X_t,Y_{t+1},Y_t) - \mathcal{H}(Y_t)
\end{equation}

The definition of the TE is asymmetric to capture the directed influence of the past of one variable on the future of another.
In the main text we use TE as a low-order counterpart to the total dynamical O-information that we use to probe higher-order behavior.
As we wish to consider the total transfer entropy present in a group of three nodes we sum up all the possible source-target combinations of the three nodes states.
Given three variables $X_1$, $X_2$, and $X_3$ this is given by: 
\begin{align}
    \label{eq:tot-TE}
    \mathcal{T}^{\rm tot.}(X_1,X_2,X_3)=&\mathcal{T}(X_1;X_2)+\mathcal{T}(X_2;X_1)+\mathcal{T}(X_1;X_3) +\nonumber\\
     & + \mathcal{T}(X_3;X_1)+\mathcal{T}(X_2;X_3)+\mathcal{T}(X_3;X_2)
\end{align}
\subsection{Total dynamical O-information}
The partial information decomposition (PID) framework is a well-established approach to characterize the information-sharing interdependencies between a group of three or more variables ~\cite{williams2010nonnegative, griffith2014quantifying, varley2023partial}.
Qualitatively, these relations can be of three types: redundant, synergistic, or unique.
Consider three variables sharing information,  $X_1$, $X_2$ and $X_3$. 
Information is said to be redundant if it is replicated over the variables (that is, recoverable from $X_1 \lor X_2 \lor X_3$), synergistic if it can only be recovered from their joint state ($X_1 \land X_2 \land X_3$), and unique if it can only be recovered from one variable and nowhere else.

In this framework, mutual information has been extended to groups of three or more variables by the so-called O-information (shorthand for ``information about Organizational structure")~\cite{rosas2019quantifying}, which on a vector of $n$ random variables $\mathbf{X}=(X_1,...,X_n)$ is given by:
\begin{equation}
    \label{eq:o_info_definition}
    \Omega_n(\mathbf{X})\equiv (n-2)\mathcal{H}(\mathbf{X})+\sum_{j=1}^n \left[\mathcal{H}(X_j)-\mathcal{H}(\mathbf{X}_{-j})\right]    ,
\end{equation}
where $\mathcal{H}(\cdot)$ is the Shannon entropy~\cite{cover1999elements} and $\mathbf{X}_{-j}=\mathbf{X}\setminus X_j $ (see ~\cite{rosas2019quantifying} for a detailed presentation of this metric's properties). 
For our purposes, the relevant property of O-information is that it is a signed metric: $\Omega_n(\mathbf{X})>0$ indicates that information-sharing is dominated by redundancy, $\Omega_n(\mathbf{X})<0$ indicates that it is dominated by synergy, and $\Omega_n(\mathbf{X})=0$ indicates a balance between both.

%dOinfo
\paragraph{Total Dynamical O-information.}
To generalize the O-information of multivariate time series from equal-time correlations to the time-lagged correlations, Stramaglia \textit{et al.} proposed \textit{dynamical} O-information $ d\Omega$ ~\cite{stramaglia2021quantifying}.
It is defined by (i) considering $n$ random variables $\mathbf{X}=(X_1,...,X_n)$ on which we have defined the standard O-information, and (ii) adding a new random variable $Y$. 
In this way, the O-information of the joint state between set $\mathbf{X}$ and the new variable $Y$ becomes:
\begin{equation}
    \label{eq:o_info_added_variable}
    \Omega_{n+1}(\mathbf{X},Y)=\Omega_{n}(\mathbf{X})+\Delta_n
\end{equation}
where:
\begin{equation}
    \label{eq:marginal}
\Delta_n = (1-n)\mathcal{I}(Y;\mathbf{X})+\sum_{j=1}^n \mathcal{I}(Y;\mathbf{X}_{-j})
\end{equation}
where $\mathcal{I}(\cdot;\cdot)$ is the canonical mutual information between two variables~\cite{cover1999elements}.
The additional term in $\Delta_n$ quantifies the variation of the total O-information induced by the addition of $Y$, effectively measuring the informational character of the circuits linking $Y$ to the variables in $\mathbf{X}$. 
In particular, if $\Delta_n$ is positive, $Y$ receives mostly redundant information from the set of variables $\mathbf{X}$, whilst if $\Delta_n$ is negative, then the influence of $\mathbf{X}$ on $Y$ is dominated by synergistic effects. 
To remove potentially confounding shared information due to common history or input signals, the dynamical O-information is defined by conditioning Eqs.~\eqref{eq:o_info_added_variable}-\eqref{eq:marginal} on the history $Y_0$ of the target variable $Y$:
\begin{align}
    \label{eq:dynamical_o_info}
    d\Omega_n(Y;\mathbf{X}) \equiv (1-n)\mathcal{I}(Y;\mathbf{X}|Y_0) + \sum_{j=1}^n \mathcal{I}(Y;\mathbf{X}_{-j}|Y_0) .
\end{align}
Here $Y_0(t)=\left(y(t),y(t-1),...,y(t-\tau+1)\right)$ and $Y=Y(t)=y(t+1)$ are the samples of $Y$ corresponding to what we consider the past and present of the variable and to its next instance, respectively.
The parameter $\tau$ is the temporal horizon of the time series and can typically be set to a relevant time scale of the process.

To quantify the dynamical information within a group of $n$ units, regardless of source-target assignments, we define the \textit{total} (or symmetrized) dynamical O-information as:
\begin{equation}
    \label{eq:total_d_o_info}
    d\Omega_n^{\text{tot.}}(\mathbf{X})\equiv\sum_{j=1}^n d\Omega_{n-1}(X_j;\mathbf{X}_{-j}) .
\end{equation}

When considering a delay $\tau=1$, as in our work, we can rewrite the dynamical O-information using the transfer entropy in the following way: 
\begin{equation}
\label{eq: Symmetrized Transfer Entropy}
    d\Omega_n(Y;\mathbf{X}) \equiv (1-n)\mathcal{T}(Y;\mathbf{X}) +\sum_{j=1}^n \mathcal{T}(Y;\mathbf{X}_{-j})
\end{equation}
In the case $\tau=1$ it is straightforward to see that the relation between the dynamical O-information and the transfer entropy is the same as the one between the O-information and the mutual information.

Using the same notation of Eq.~\eqref{eq:Transfer Entropy} we can specialize Eq.~\eqref{eq:dynamical_o_info} to the case $\tau=1$ as:
\begin{equation}
    d\Omega_n(Y;\mathbf{X}) \equiv (1-n)\mathcal{I}(Y_{t+1};\mathbf{X}_t|Y_t) + \sum_{j=1}^n \mathcal{I}(Y_t;\mathbf{X}_{-j,t}|Y_{t}) .
\end{equation}
where we see that conditioning on the past ($Y_t$) is applied in the same way to the two measures.

\subsection{Computation of information theoretical metrics}
The computation of the information theoretical metrics Eq.~\eqref{eq:tot-TE} and Eq.~\eqref{eq:total_d_o_info} is based on the computation of the entropy of a random variable.
In the two systems considered in our work the random variables under consideration---\textit{i.e.} the state of the nodes in the simplicial complex---are binary.
Given a set of $M$ observations $\{x_i\}\in [0,1]^N$ of the state of a binary random variable $X$ taking values $x\in [0,1]$, we compute the empirical entropy of $X$ as:
\begin{equation}
\label{eq:empirical_binary_entropy}
    \mathcal{H}(X)=-\sum_{x} p_x \log p_x
\end{equation}
where:
\begin{equation}
    p_x = \frac{1}{M}\sum_{i=1}^M \delta_{x_i,x}
\end{equation}
is the observed frequency for $x$.
The extension of Eq.~\eqref{eq:empirical_binary_entropy} to the multidimensional random variables is straightforward.

The computational tools developed in this work to compute the information theoretical metrics are now part of the publicly available HOI Python package (\url{https://brainets.github.io/hoi/}).
The HOI package uses high-performance numerical computing tools to mitigate the high computational costs involved in exploring higher-order interdependencies.

\section{Higher-order dynamical models}

\subsection{Simplicial Ising model}
\paragraph{Low-order Ising model.}
The Ising model is a mathematical model of ferromagnetisms in statistical physics~\cite{huang2009introduction}. 
It was originally defined on lattices and later extended to complex networks.
We consider the general case of the model being defined on a graph $\mathcal{G}=(V,E)$, with $N=|V|$ nodes.
To each node $i \in V$ we associate a binary state variable $S^i\in \{\pm 1\}$, corresponding to the spin state of the node, which can either be up ($S^i=+1$) or down ($S^i=-1$).
The global state of the system is controlled by the Hamiltonian:
\begin{equation}
\label{eq:ising_hamiltonian}
    H=-\sum_i h_iS^i -\sum_{\langle i,j\rangle} J_{ij} S^i S^j
\end{equation}
where $\{h_i\}$ are the local magnetic fields and $\{J_{ij}\}$ are the magnetic couplings for the neighbouring nodes $\{\langle i,j\rangle\}$.
A special case of Eq.~\eqref{eq:ising_hamiltonian} that allows for an exact analytical treatment in one and two-dimensional systems and a simple mean-field formulation is obtained with no magnetic fields (\textit{i.e.} $h_i=0 \; \forall i \in V$) and a positive uniform magnetic coupling $J_{ij}=J \;\forall \langle i,j\rangle \in \mathcal{G}$ with $J>0$.
In this case, the Hamiltonian of the systems takes the form:
\begin{equation}
\label{eq:uniform_ising_hamiltonian}
H=-J\sum_{\langle i,j\rangle} S^i S^j
\end{equation}
The energy of the system described by Eq.~\eqref{eq:uniform_ising_hamiltonian} is minimized when all spins are aligned.
The order parameter of these Ising models is the magnetization:
\begin{equation}
\label{eq:magnetization}
m=\frac{1}{N}\sum_{i} S^i
\end{equation}
This parameter takes values in the interval $[-1,+1]$.
The system described by Eq.~\eqref{eq:uniform_ising_hamiltonian} undergoes a second-order phase transition between a fully disordered state ($m=0$) and magnetized state ($|m|=1$) at a critical value of the temperature $T$.

\paragraph{Extension to higher-order.}
The extension of the Ising model to incorporate many-body interactions is not a recent question in statistical physics~\cite{baxter1973exact, merlini1973symmetry, kirkpatrick1987dynamics}. 
Notably, so-called \textit{plaquette} interactions involving groups of 4, 8, 16, \textit{etc.} nodes appear when performing real-space renormalization on a square lattice, and higher-order terms appear when performing other renormalization procedures~\cite{goldenfeld2018lectures}.
Moreover, the $p$-spin model~\cite{derrida1980random}, a generalization of the disordered Ising model given by Eq.~\eqref{eq:ising_hamiltonian} with infinite range interactions between groups of spins of size $p$, is extensively studied in the spin glass literature.
Recently, a direct extension of Eq.~\eqref{eq:uniform_ising_hamiltonian} has been proposed to describe an Ising model on a simplicial complex of order $\ell=2$~\cite{wang2022full}. 
This model has a Hamiltonian given by:
\begin{equation}
\label{eq:three_body_ising}
    H = -J_0\sum_{\langle i,j\rangle} S^iS^j -J_1\sum_{\langle i,j,k\rangle} S^iS^jS^k
\end{equation}
where $J_1$ and $J_2$ are the strengths of two-body and three-body interactions, and $\langle i,j\rangle$ and $\langle i,j,k\rangle$ denote the two-body and the three-body connections in the 2-simplicial complex, respectively.
However, as was already noted in the first analysis of the Ising model with three-body interactions, the extension provided by Eq.~\eqref{eq:three_body_ising} breaks the parity symmetry under spin flip at all sites of the dyadic model without magnetic field~\cite{merlini1973symmetry}.
This can be easily understood considering the energy of systems constituted by three spins $S^1$, $S^2$, and $S^3$ connected in a 2-simplex. 
There are two states with all spins aligned: all spins pointing up and all spins pointing down.
These two configurations are symmetric upon flipping all spins, yet their energies when computed using Eq.~\eqref{eq:three_body_ising} are different, favoring the state with spins pointing up.
To preserve the symmetry of the original low-order Ising model we propose an alternative formulation of the extension of the Ising model to simplicial complexes of arbitrary order.
For every simplex $\sigma$---of all orders---we consider its only configuration favored in energy to be the one with all the spins aligned.
This leads to formulating a new Hamiltonian for the simplicial Ising model:
\begin{equation}
\label{eq:simplicial_ising_correct}
    H = - J_0 \sum_{i=1}^N S^i
    - \sum_{\ell=1}^{\ell_{\text{max}}} J_{\ell}
    \sum_{\{\sigma \in \mathcal{K}:|\sigma|=\ell\}}
    \left[2 \bigotimes_{i\in \sigma}S^i -1 \right]
\end{equation}
where:
\begin{equation}
    \bigotimes_{i=1}^n S^i=\delta\left(S^1,...,S^n\right) =\begin{cases}
    1 & \text{if\;} S^1=S^2=...=S^n \\
    0 & \text{otherwise}
    \end{cases}
\end{equation}
is the Kronecker delta with an arbitrary number of binary arguments and $\ell_{\rm max}$ is the maximal order of $\mathcal{K}$.
For $\ell_{\rm max}=1$ Eq.~\eqref{eq:simplicial_ising_correct} reduces to the low-order model with uniform magnetic field $J_0$ and pairwise coupling $J_1$.
Upon rescaling the couplings by the corresponding generalized degrees, we obtain the Hamiltonian used in our work.

\paragraph{Monte Carlo dynamics.}
We consider the dynamics of this system to be the Markov chain of Monte Carlo moves performed with Metropolis-Hastings acceptance-rejection rule~\cite{newman1999monte} at temperature $T$.
Monte Carlo dynamics are not defined to be dynamical models but rather numerical methods for solving statistical physics problems.
However, the Metropolis-Hastings acceptance-rejection rule defines Markovian transitions between configurations of the model which we consider---to our scope---to be the dynamic of the system.
We start from a random configuration in which each spin has equal probability of being in one of the two states.
At each time step $t$ we randomly select a set of independent spins (\textit{i.e.} sites in the simplicial complex $\mathcal{K}$ that are not first-neighbors).
For each selected spin we propose a flipping move.
In the case considered in the main text---simplicial complex with $\ell_{\rm max}=2$ and no magnetic field, $J_0=0$---the energy variation given by the flipping of spin $i$ is given by:
\begin{equation}
    \Delta E^i(t)= 2 \frac{J_1}{\langle k_1\rangle} \sum_{j\in \partial_i} \left[2\delta\left(S^i\left(t\right), S^j\left(t\right)\right)-1\right] + 
    2 \frac{J_2}{\langle k_2\rangle} \sum_{(j,k)\in \nabla_i} \left[2\delta\left(S^i\left(t\right), S^j\left(t\right),S^k\left(t\right)\right)-1\right]\delta \left(S^j(t),S^k(t)\right)
\end{equation}
where $\partial_i$ is the set of nodes in $\mathcal{K}$ connected to node $i$ by an edge and $\nabla_i$ is the set of pairs of nodes in $\mathcal{K}$ forming a 2-simplex with node $i$.
The acceptance or rejection of the proposed move is based on the Metropolis-Hastings rule, with the acceptance probability given by:
\begin{equation}
    P\left(S^i(t) \to -S^i(t)\right)=
    \begin{cases}
        \exp\left[-\frac{\Delta E^i(t)}{T}\right] & \text{\;if\;} \Delta E^i(t) >0\\
        1  & \text{\;otherwise}
    \end{cases}
\end{equation}
In other words, if we select a new state that has an energy lower than or equal to the present one, we should always accept the transition to that state.

\subsection{Simplicial contagion model}
The simplicial contagion model~\cite{iacopini2019simplicial} is a dynamical model that describes social contagion processes such as opinion formation or the adoption of novelties, where complex mechanisms of influence and reinforcement are at work.
The model is defined on a simplicial complex $\mathcal{K}$ with $N$ nodes.
In this model, the standard susceptible-infected-susceptible (SIS) compartmental model for contagion processes~\cite{vespignani2012modelling} is extended to account for group interactions.
Following the SIS framework~\cite{pastor2015epidemic}, to each node of a simplicial complex $\mathcal{K}$ we associate a binary random variable $x_i(t)\in \{0,1\}$ such that at each time step we divide the population of individuals into two classes of susceptible (S) and infectious (I) nodes, corresponding respectively to the values 0 and 1 of the state variables $x_i(t)$.
At the initial time step a finite fraction of infected agents $\rho_0$ is placed in the population.
At each time step, each susceptible agent ($x_i(t)=0$) becomes infected with a probability $\beta_{\ell}$ if it belongs to a $\ell$-simplex where all the other $\ell$ nodes are infected.
The infected agents recover independently with probability  $\mu$.
The order parameter of this model is the density of infected agents at time $t$, given by:
\begin{equation}
\label{eq:density-of-infected}
\rho(t)=\frac{1}{N}\sum_i x_i(t)
\end{equation}

\paragraph{Rescaled infectivity rates.}
In the original paper~\cite{iacopini2019simplicial}, the authors provide an analytical characterization of the mean-field description of this model on simplicial complexes with $\ell_{\rm max}=2$. 
Given the set of infection probabilities $\{\beta_{\ell}\}$ and the recovery rate $\mu$, assuming homogeneous mixing and the independence between the states of different nodes, the mean-field expression for the temporal evolution of the density of infected nodes $\rho(t)$ is:
\begin{equation}
    \dot{\rho}(t)=-\mu \rho(t)+\sum_{\ell=1}^{\ell_{\rm max}}\beta_{\ell}\langle k_{\ell}\rangle \rho^{\ell}(t)[1-\rho(t)]
\end{equation}
This expression can be simplified and treated analytically by rescaling time by a factor $\mu$ and introducing the rescaled infectivity rates $\{\lambda_{\ell}=\mu\beta_{\ell}/\langle k_{\ell}\rangle\}$.
These rescaled infectivity rates are the control parameters we consider in our work.

\section{Numerical values of the total dynamical O-info}

\begin{figure}
    \centering  
    \includegraphics[width=0.9\textwidth]{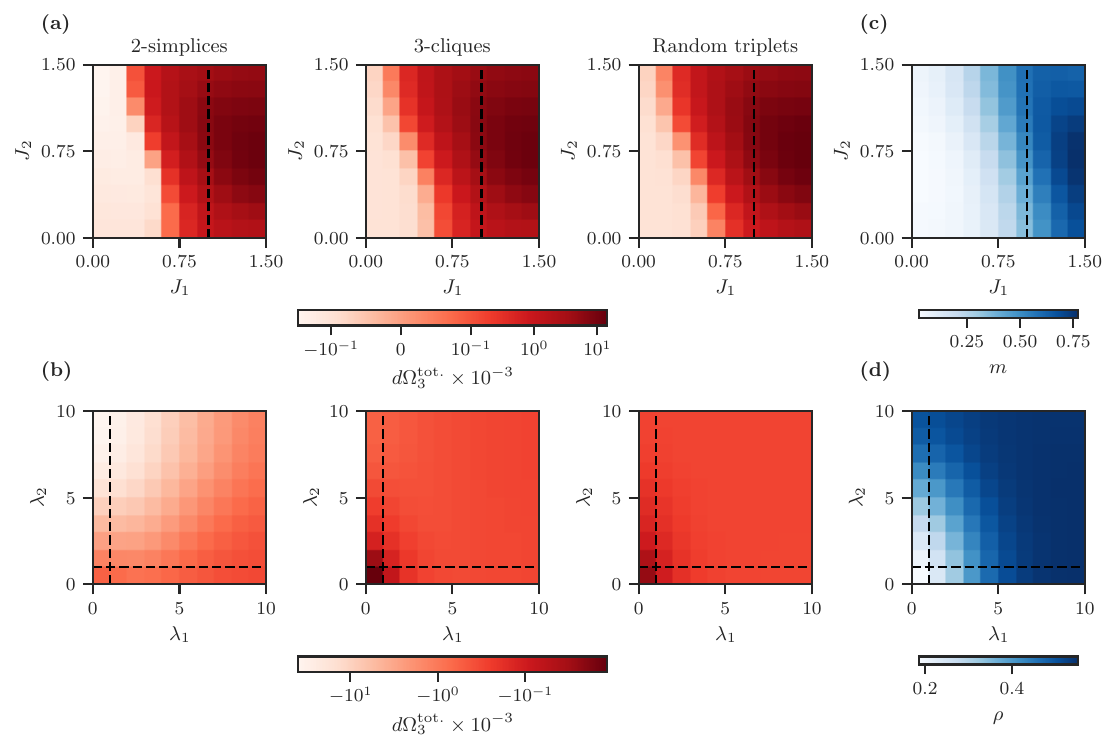}
    \caption{
    (a)-(b)~Numerical values of the $d\Omega_3^{\rm tot.}$.
    (c)-(d)~Order parameters (magnetization and density of infected agents).
    (a)~Ising model.
    Results are obtained on a random simplicial complex with $N=200$ nodes and average degrees $\langle k_1 \rangle = 20$, $\langle k_2 \rangle = 6$, running the Monte Carlo dynamics for $3\times 10^4$ time steps with $J_0=0$ at temperature $T=1$. 
    The dashed line is the critical coupling strength of the pairwise model $J_1^{\rm cr.}=1$ with no magnetic field. 
    (b)~Simplicial contagion model. 
    Results are obtained on a random simplicial complex with $N=200$ nodes and average degrees $\langle k_1 \rangle = 20$, $\langle k_2 \rangle = 6$, running the contagion dynamic for $10^4$ time steps with $\mu = 0.8$ and $\rho_0=0.3$.
    The two dashed lines are respectively the epidemic threshold of the pairwise SIS model $\lambda_1^{\rm cr.}=1$ and the critical value of the rescaled 2-simplices infectivity rate above which the system shows the discontinuous phase transition and bistability $\lambda_2^{\rm cr.}=1$.}
    \label{fig:abs-doinfo}
\end{figure}

In the main text, we focus on the statistical distance between the distributions of the total dynamical O-info of 2-simplices and 3-cliques. 
In Fig.~\ref{fig:abs-doinfo}a,b we show the numerical values of $d\Omega_3^{\rm tot.}$ in the parameter space.
For additional reference, we also show in Fig.~\ref{fig:abs-doinfo}c,d the numerical values of the order parameter of the two models (see Eq.~\eqref{eq:magnetization} and Eq.~\eqref{eq:density-of-infected}).
In the Ising model (panels a,c), we see that in the phase in which the system is not magnetized all groups of three nodes (2-simplices, 3-cliques, and random triplets) behave synergistically, with the 2-simplices showing more negative values of $d\Omega_3^{\rm tot.}$. 
As the system magnetizes---this is particularly evident above the threshold of the pairwise model---information is duplicated across units of the system and thus the groups of nodes share redundant information, $d\Omega_3^{\rm tot.}\geq 0$.
In the simplicial contagion model (panels b,d), we see that overall the groups of three nodes in the system behave synergistically with the 2-simplices always displaying more negative values of $d\Omega_3^{\rm tot.}$ with respect to the other groups of three nodes.
The reason for the fact that in the simplicial contagion model groups don't behave redundantly---as instead happens in the Ising model---when the order parameter is large (\textit{i.e.} in the endemic phase of the contagion model) is because infected nodes recover independently one from the other and independently from the global state of the system. 
For this reason, even in the endemic phase, the state of nodes is not frozen as happens in the magnetized phase of the Ising model and thus the synergistic behavior dominates over the redundant contribution and the $d\Omega_3^{\rm tot.}$ remains negative.

\begin{figure}
    \centering
    \includegraphics{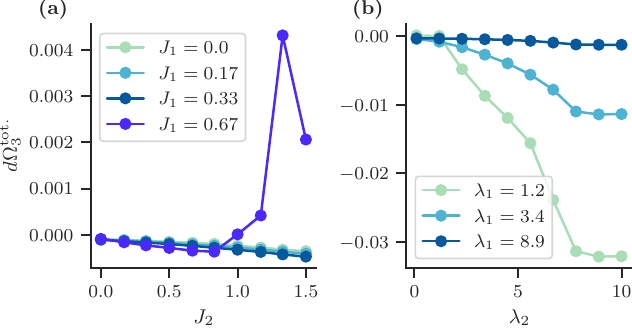}
    \caption{Non-linearity of the relation between the total dynamical O-information of 2-simplices and the strength of the parameters controlling the group interactions ($J_2$ and $\lambda_2$ respectively).
    (a)~Ising model. (b)~Simplicial contagion model.}
    \label{fig:non-linear}
\end{figure}

We can explicitly see the non-linearity of the relation between higher-order behaviors and higher-order mechanisms---mentioned in the main text when looking at the dependency of the relative higher-order behavior on the strength of the parameters controlling the group interactions---in Fig.~\ref{fig:non-linear}.
In the figure, we show the dependence of the total dynamical O-information of 2-simplices and the strength of the parameters controlling the group interactions ($J_2$ and $\lambda_2$ respectively).
The jump in the Ising model for $J_1=0.67$ is due to the transition to the magnetized phase, resulting in a redundancy-dominated interdependency between the units of the 2-simplices.

{}

\section{Differences between the behavior of 3-cliques and random triplets}
\begin{figure}
    \centering
    \includegraphics{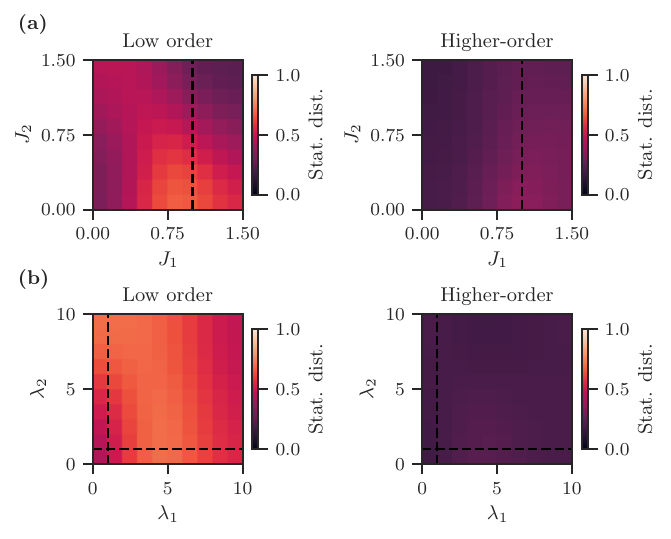}
    \caption{
    \textbf{
    Statistical distance between the distributions of low (sum of transfer entropies) and higher-order (total dynamical O-information) observables of behavior of 3-cliques and random triplets.}
    (a)~Ising model.
    (b)~Simplicial contagion model.}
    \label{fig:distance_cliques_random}
\end{figure}

In the main text, we focus mainly on the difference measured by the information-theoretical metrics for behavior between 2-simplices and 3-cliques.
In Fig.~\ref{fig:distance_cliques_random} we show the statistical distance between the distributions of low and higher-order observables of behavior between 3-cliques and random triplets (\textit{i.e.} groups of three nodes not connected by a 2-simplex or a 3-clique).
We see that overall the sum of transfer entropies provides a better discrimination between 2-cliques and random triplets with respect to total dynamical information.
This is a comforting result for two reasons: (i) low-order observables provide some discrimination between the interactions in the system and random group of nodes and (ii) as observed in the main text from the higher-order point of view 3-cliques and random triplets show similar behaviors.
In the Ising model (panel a) we see that the statistical distance for the low-order observables is larger in the vicinity of the critical point of the pairwise system.
This echoes known results about the inverse Ising problem on graphs allowing for solutions based on mean-field inverse correlations near the critical point~\cite{nguyen2017inverse}.
In the simplicial contagion model (panel b) the statistical distance between the sum of transfer entropies in 3-cliques and random triplets is large in the majority of the space of the parameters.
There are two regions in which the statistical distance is slightly smaller: for small values of both $\lambda_1$ and $\lambda_2$ (bottom left corner) and for large values of both the parameters. 
The reduced statistical distance in these two regions can be understood considering the dynamic of the system. 
When $\lambda_1$ and $\lambda_2$ are small the number of infected agents in the system is small or null, thus the state of nodes does not change much in time and we see no difference between interacting and disconnected nodes.
Likewise, when $\lambda_1,\lambda_2 \gg 1$ the majority of the nodes are infected and thus also the states of non-interacting nodes are more correlated.

\section{Synergy-based detection of higher-order interactions}
We can exploit the relation between higher-order mechanisms and behaviors to detect interactions from dynamic.
We set ourselves in the case in which we want to discriminate between true higher-order interactions (2-simplices) and spurious ones (3-cliques) having access to the nodes' states' time series, the low-order structure of the system and its average generalized degree of order $\ell=2$.
This setup reproduces the situation in which we have access to the node-level activity of a system, its pairwise structure, and some coarse information about nodes' neighborhoods in terms of their higher-order connectivity (\textit{e.g.} through local samples).

The heuristic method proceeds as follows.
Given  $\langle k_2\rangle$ and the number of nodes $N$, the expected number of 2-simplices in the system is $n_{\triangle}=N\langle k_2 \rangle/3$.
Given $n_{\triangle}$, we compute the total dynamical O-information for all triplet groups connected by pairwise connections (\textit{i.e.} the 3-cliques in the skeleton of the simplicial complex $\mathcal{K}$ under study), rank them by their values of $d\Omega_3^{\rm tot.}$ and mark as ``predicted" higher-order interactions the $n_{\triangle}$ most synergistic ones.

\begin{figure}
    \centering
    \includegraphics{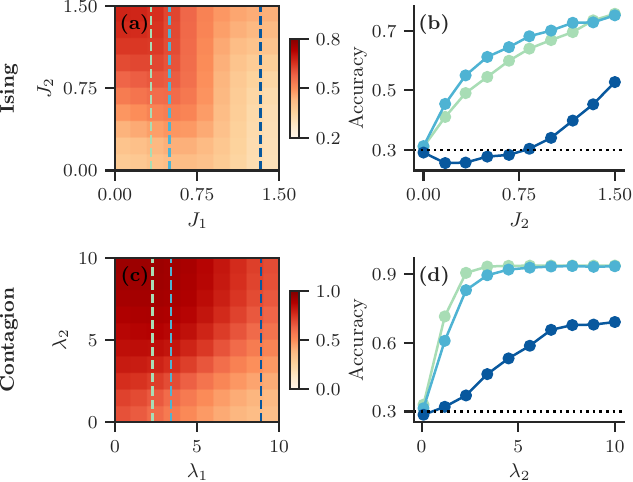}
    \caption{
    \textbf{
    Detection of higher-order interactions using higher-order synergistic behavior.}
    We show the accuracy (fraction of true positives) of the detection of higher-order interactions for (a)-(b)~the Ising and 
    (c)-(d)~the simplicial contagion models. 
    (a), (c) Full parameter space, and (b), (d) show three example curves corresponding to the dashed lines in (a) and (c). The dotted lines in (b) and (d) indicate the random choice accuracy.
    }
    \label{fig:phasespace_score}
    \vspace{-1em}
\end{figure}

Fig.~\ref{fig:phasespace_score} shows the results of the method in terms of accuracy scores, which
retrace our previous observations based on the different higher-order behaviors of 2-simplices and 3-cliques.
While reconstruction scores are larger for the contagion (accuracy $\sim 0.9$) than for the Ising model (accuracy $\sim 0.7$), in the region of synergistic signatures they are always significantly better than random choice accuracy ($0.3$, see below).
These scores show also a complex non-linear dependence on the higher-order mechanisms' strength (Fig.~\ref{fig:phasespace_score}b,d), which is a byproduct of the non-linear response of higher-order behaviors to the variation of higher-order mechanisms.
Applying this method to more general set-ups (\textit{e.g.} group interactions among all possible combinations of three nodes) leads to similar results.

\subsection{Random choice accuracy}
The random choice accuracy score in our detection method is given by the ratio between $n_{\triangle}$ and the number of 3-cliques in the skeleton of the simplicial complex under study.
The nominator of this ratio is computed using the generalized degree as:

\begin{equation}
\label{eq:number-of-simplices}
    n_{\triangle}=\frac{N\times \langle k_2 \rangle}{3}
\end{equation}

In the systems used for our results, the denominator in the accuracy ratio can be easily computed, as the skeleton of a random simplicial complex is an Erdős–Rényi (ER) random graph.
The number of 3-cliques in a ER random graph with $N$ nodes and edge probability $p$ (and average degree $\langle k_1 \rangle=\langle k \rangle=p(N-1)$) is given by:
\begin{equation}
\label{eq:number-of-cliques}
    \mathbb{E}_{N,p}[\# \text{\;of 3-cliques}]=\binom{N}{3}p^3
    =\frac{N(N-2)\langle k\rangle^3}{6(N-1)^2}
\end{equation}
Using the parameters employed in our simulations ($N=200$, $\langle k_1\rangle = 20$, and $\langle k_2 \rangle=6$), taking the ratio between Eq.~\eqref{eq:number-of-simplices} and Eq.~\eqref{eq:number-of-cliques} we obtain a random choice accuracy of 0.3.

\bibliography{reference_supp}